\documentclass[sigconf,nonacm]{acmart}
\usepackage{tikz}

\AtBeginDocument{%
  \providecommand\BibTeX{{%
    \normalfont B\kern-0.5em{\scshape i\kern-0.25em b}\kern-0.8em\TeX}}}

\makeatletter
\def\@ACM@checkaffil{
    \if@ACM@instpresent\else
    \ClassWarningNoLine{\@classname}{No institution present for an affiliation}%
    \fi
    \if@ACM@citypresent\else
    \ClassWarningNoLine{\@classname}{No city present for an affiliation}%
    \fi
    \if@ACM@countrypresent\else
        \ClassWarningNoLine{\@classname}{No country present for an affiliation}%
    \fi
}
\makeatother

\usepackage{xcolor}

\usepackage[acronym]{glossaries}
\newacronym[longplural=Software Bills of Materials, shortplural=SBOMs]{sbom}{SBOM}{Software Bill of Materials}
\newacronym{vex}{VEX}{Vulnerability Exploitability eXchange}
\newacronym{s3c2}{S3C2}{Secure Software Supply Chain Center}
\newacronym{nsf}{NSF}{National Science Foundation}

\usepackage[style=american,threshold=2,autopunct]{csquotes}

\renewcommand{\mkbegdispquote}[2]{\leavevmode\llap{``}}

\usepackage{tcolorbox}
\tcbuselibrary{breakable}
\tcbuselibrary{skins}
\definecolor{darkgray}{gray}{0.3}
\tcbset{}
\newtcolorbox{summaryBox}[2][]
{
	enhanced,
	breakable,
	frame hidden,
	borderline west = {2pt}{0pt}{lightgray},
	colback         = white,
	size            = fbox,
	left            = 0.5em,
	coltitle        = black,
	title           = {\color{darkgray}#2. },
	attach title to upper,
	#1,
}

\begin{document}

\title{S3C2 Summit 2024-03: \\ Industry Secure Supply Chain Summit}

\author{
Greg Tystahl$^{\ddagger}$,
Yasemin Acar$^{*}$,
Michel Cukier$^{\dagger}$,
William Enck$^{\ddagger}$,\\
Christian Kästner$^{\mathsection}$,
Alexandros Kapravelos$^{\ddagger}$,
Dominik Wermke$^{\ddagger}$,
Laurie Williams$^{\ddagger}$}

\def \authors{%
Greg Tystahl,
Yasemin Acar,
Michel Cukier,
William Enck,
Christian Kästner,
Alexandros Kapravelos,
Dominik Wermke,
Laurie Williams}

\affiliation{%
    \vspace{.5em} 
    \institution{ $^\ddagger$North Carolina State University, Raleigh, NC, USA}
}
\affiliation{%
    \institution{$^*$Paderborn University, Paderborn, Germany, and George Washington University, DC, USA}
}
\affiliation{%
    \institution{$^\dagger$University of Maryland, College Park, MD, USA}
}
\affiliation{%
    \institution{ $^\mathsection$Carnegie Mellon University, Pittsburgh, PA, USA}
}

\renewcommand{\shortauthors}{Secure Software Supply Chain Center (S3C2)}
\renewcommand{\shorttitle}{S3C2 Summit 2024-03: Industry Secure Supply Chain Summit}

\begin{abstract}
Supply chain security has become a very important vector to consider when defending against adversary attacks. 
Due to this, more and more developers are keen on improving their supply chains to make them more robust against future threats. 
On March 7th, 2024 researchers from the Secure Software Supply Chain Center (S3C2) gathered 14 industry leaders, developers and consumers of the open source ecosystem to discuss the state of supply chain security.
The goal of the summit is to share insights between companies and developers alike to foster new collaborations and ideas moving forward. 
Through this meeting, participants were questions on best practices and thoughts how to improve things for the future. 
In this paper we summarize the responses and discussions of the summit. The panel questions can be found in the appendix.
\end{abstract}

\settopmatter{printfolios=true}

\maketitle

\begin{tikzpicture}[overlay, remember picture]
\node[anchor=north west, 
      xshift=17.5cm, 
      yshift=-2.1cm] 
     at (current page.north west) 
     {\includegraphics[width=2.1cm]{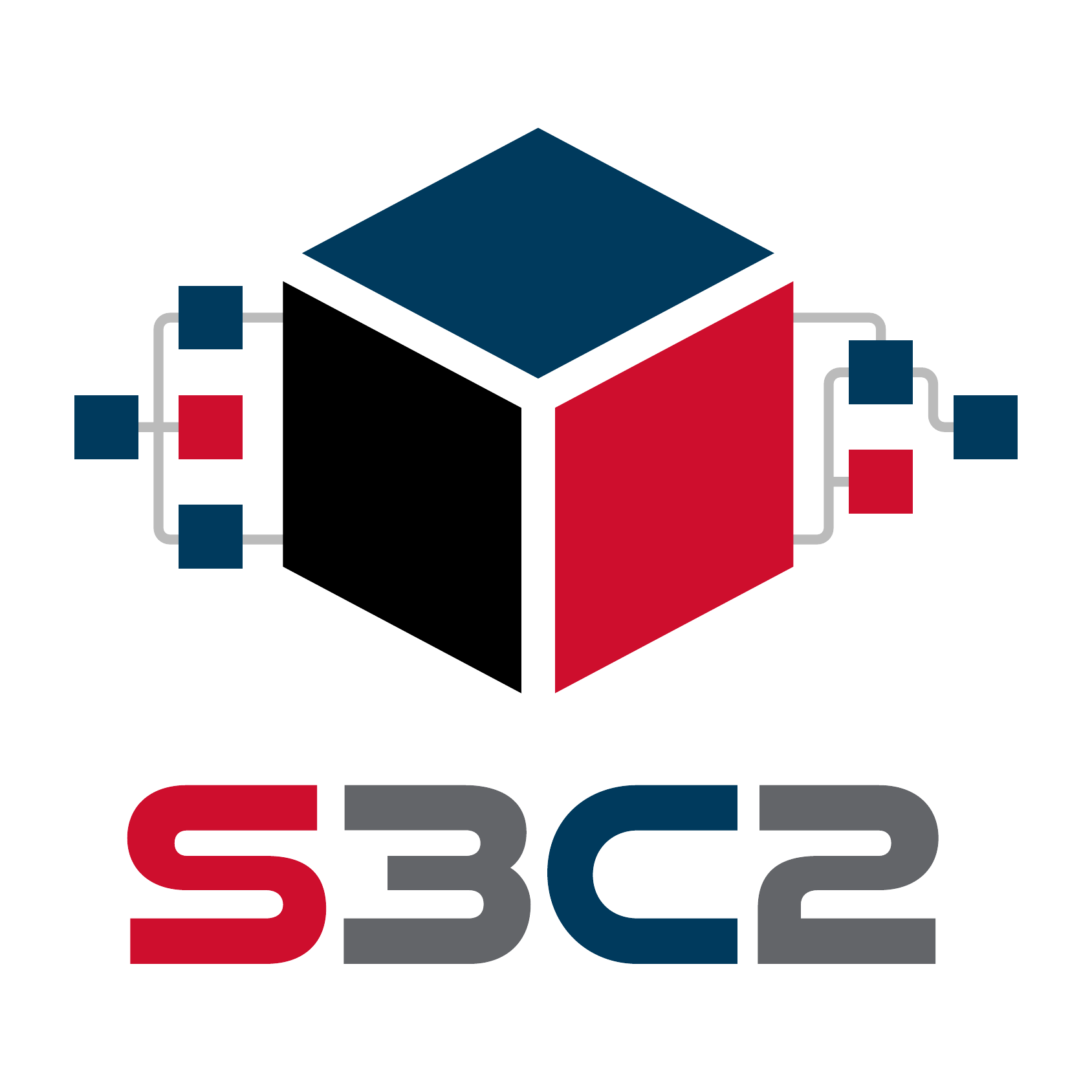}}; 
\end{tikzpicture}

\section{Introduction}
Supply Chain security has become a major talking point of industry practitioners.
From the ashes of the Spectre, SolarWinds, and Log4J attacks came the executive order requiring a higher focus on supply chain security \cite{eo:2021:improving}. 
Through this order, work has been done to improve software visibility through SBOMs \cite{zahan2023software}, software attestations, and build frameworks \cite{SLSA}. 

As software grows, policies may need to be reevaluated to capture all potential scopes and emerging areas of concern.
With the explosive growth in AI advancements as well as wide spread adoption of large language models (LLMs), areas of the supply chain may gain new attack vectors that need to be considered.

To this end, a group of researchers from the Secure Software Supply Chain Center gathered 14 industry leaders, developers, and consumers of the open source ecosystem to discuss the state of supply chain security. 
This summit was held on March 7th, 2024 and lasted for one day. 
Participants of the summit were questioned in six different panels by other participants. 
The summit followed Chatham House Rules where participants were allowed to talk about what was discussed in the summit, but were not able to reveal who said what.
As such, information disclosed in this paper disseminating what was discussed will not include any information as to who participated in this meeting. 

The next sections are formatted based on the panel questions and made up of summarized responses from the entire group.
A full list of panel questions can be found in the appendix. 

\section{Software Bill of Materials (SBOM)}

SBOMs have gathered a large popularity due their requirement in the executive order \cite{eo:2021:improving}. 
Generation of SBOMs has become a common practice of producers of software using tools such as CycloneDX and SPDX.
While the adoption of SBOMs have increased, it is unclear as to how companies are using the SBOMs that are created.

\subsection{SBOM Creation}
While creation of SBOMs may have been a huge limiting factor in the past, SBOM creation has become much easier. 
CycloneDX and SPDX are the two major generators used by the participants. 
These tools have become mostly interchangeable as their formats are largely dependent on the specifications of the executive order.

However, while the generation process has become easier, the challenge still remains for how many SBOMs need to be created. 
Some of the participants debated on how much of their software should have SBOMs generated.
For software that has multiple builds dependent on the operating system that they are building for, should there be one major SBOM for all possible builds types or multiple per build?
The answer gathered was dependent on the consumer of the SBOM. 
If requested, they would generate multiple SBOMs if needed.

\subsection{SBOM Consumption}
The general consensus was that no consumer knows what to do with SBOMs once they get them. 
A lot of consumers are asking for SBOMs as part of the compliance, but they are not using them for anything at this time as \textbf{use} is not required.

While consumers may not be using SBOMs, one of the participants discussed the internal use case for their generated SBOMs.
They created their own analysis software that uses SBOMs for easier software composition analysis (SCA).
With the use of SBOMs, a full SCA is not needed every time a new version of something comes out. Using the SBOM to find just the components that need to change seemed to be a big help.
Another participant discussed the idea that SBOMs are making it easier for companies to communicate their software to each other.
They stated that SBOMs themselves may not be the best solution, but they are getting companies to start understand supply chain security.
Understanding what software is being added to dependent software has made it easier to track down potential risks.
To this effect, one participant stated that this kind of visibility will lead to companies being embarrassed about using very out of date or vulnerable software and will increase security updates accordingly to avoid this.

\subsection{VEX Use}
Creating VEX statements is not possible for all scenarios with SBOMs.
Since VEXs are dynamic, they need to be generated continuously for a specific SBOM, which are statically created.
Work needs to be done to be able to create more lightweight VEX statements to accommodate multiple SBOMs at the same time.
Otherwise, participants believed it to be infeasible to create VEX statements for every SBOM generated as well as inefficient.

\subsection{Challenges}
While SBOMs have seem improvments, participants still faced some challenges as to their adoption. 
One of the challenges that they faced  was how effective are they at increasing security? 
It can be said that knowing what components are in dependent software is important, but is SBOM giving any more information that cannot be gathered by other means?
SCA tools have existed for quite some time without the existence of SBOMs, so are they truly needed or do SCA tools need to adopt similar functionality?
Work needs to be done to prove just how SBOMs themselves reduce attack vectors for their cost to increase their use.

\section{Vulnerable Dependencies}\label{vulnDep}

Dependency management is an integral part of the supply chain, especially when it comes to vulnerable dependencies. 
Understanding which components to use as a dependent, detecting if they are vulnerable, and how to fix found vulnerabilities or replace them with similar non-vulnerable counterparts are all important things to consider.
We asked participants to talk about their processes for this in this panel.

\subsection{Dependency Choice}
There were multiple different answers for how to choose which dependencies to use. 
Some of these answers depend on the size of the company and how much available resources they are able to allocate to this.
One participant described their heavyweight processes to check the dependencies before implementing them into the code.
If any CVS found within the component have a score of 7 or higher, the component is not cleared for use.
While this is able to capture a lot of problems before they are implemented, this is a large effort.

Another way to discover potential problems was building all software in house.
Building in house allows for using the same scans as normal source code to find any vulnerabilities.
However, this may not be possible for all kinds of dependencies where source code is not available. 

\subsection{Dependency Update Policies}
Updating dependencies that are either vulnerable or out of date requires as large amount of effort. 
A participant stated some companies have no way to stay up to date with all of their dependencies for every version that is released (minor or major).
On top of this, they also do not have a way to come up with a total cost for an update.

Thus, a participant proposed their solution which is having a policy that fits the level behind you are willing to stay. 
Their example policy that has worked well for the most part is to be behind the release by three minor releases.
Another example policy was doing patches quarterly and only updating outside of this for CVE releases.
This increases security over not updating at all while still reducing the total effort required. 

One point of discussion was the idea of patching fatigue.
Having to update for every version may lead to burning out the developers.
On top of this, it was discussed that the latest patch may not be the most secure one.
To fix fatigue, another solution proposed was automating the updates.
It was something that split the participants. 
Some stated that they leverage automatic tools like Dependabot to handle this, but believe it is not fully there yet.
Automatic updates, while they have gone a long way tends to adds too much noise and overhead on things that break to be fully effective currently.

\section{Malicious Commits}

Due to the nature of open source software, there is always the lingering possibility that an attacker can commit seemingly benign malicious code into software products. 
A recent example of this was the XZ Backdoor \cite{XZLessons}.
In this attack a malicious actor gained co-control of a low level software repository and gained remote code execution on a large number of seemingly secure devices.
We asked the panel how they are able to detect such commits and how the ecosystem as a whole can get better at detecting them.

\subsection{Detection}
Participants in this study had mixed answers on how to truly detect malicious commits.

One half of them stated that there is no way to detect malicious commits automatically.
They believe that malicious commit detection is an unsolvable problem and to fix this must be done in the environment and deployments rather than in the code creation.

The other half believed that developers can follow the signals that have been extracted in the past such as behavior.
The rebuttal to this was that automatically detecting new, non-trivial commits is seemingly impossible.
At this time nobody knows exactly how to define truly what is truly considered a malicious signal. The big problem with detection of maliciousness was that the line between a mistake and maliciousness is too close.
Looking for behavior is a signal, but it does not irrefutably mark a commit as malicious.
Non trivial examples blur the line even further. 
Third-party dependencies that are added by a malicious actor with intent to be changed later can still be malicious but not look that way on the first commit.

Reputation was brought up as an important vector to consider because it can be both a good and bad pattern for malicious commits.
Attacking the accounts that have the good reputation in the open source ecosystem to use allows for attackers to gain control much faster than if they created their own new accounts.
This type of attack is very hard for a human to detect and relies heavily on the account security of the popular user.

\subsection{Remediation}
While malicious commits can be a huge problem, participants debated potential strategies to remediate the impact of them.
One of the participants provided code reviews done by the developers or outside parties help with catching malicious commits, but they do not capture everything.
In really small projects with only one developer, who will or should do the code review?
This becomes especially important if the main developer themselves becomes malicious. 

Another participant suggested signing as a way to track the commits back to the committer with a very hard forgability. 
This does not stop malicious commits from occurring, but it can make it easier to remove the malicious user or patch their breach once it is found.
Attestation, like what is required of SLSA Build Level 2 falls into a similar vein \cite{SLSA}.
While this is able to show a trail of where commits or vulnerable projects come from, it does not prevent nor detect malicious commits from happening.

A major remediation strategy proposed and almost universally accepted was build speed.
Being able to respond quickly to the detection of vulnerabilities added through a malicious commit was almost more important than detection beforehand.
The caveat to this was that faster builds requires very stable and good software engineering practices to respond quickly.

\subsection{Malicious Components VS Malicious Commits}
There was a clear sentiment among the participants that there needs to be made a difference between malicious commits and malicious components.
Understanding what the component should do as normal behavior and then detecting anomalies is easy.

Behavior analysis of the components is more powerful than of the commits themselves and can be used to capture a lot of maliciousness.
This kind of analysis captures attacks such as typo squatting or complete takeover of a dependency.
This works for dependencies that you can analyze as well as ones that are harder like Docker.
A factor in which this kind of analysis becomes difficult is for new software. 
It is hard to tell what is normal for very new software.
To make this even stronger and more automatic, it was suggested to use machine learning to determine what is the normal and to detect the abnormal.

Most of these strategies proposed for both commits and components work for third party and internal code.

\section{Build Infrastructure}

Understanding what should be done to secure the build and deploy process pipeline depends heavily on the company.
If the company has a more strict risk policy then this will affect the pieces that go into their builds.
It also depends heavily on the build environment that the build is being done in.
If building on GitHub, the SLSA generation tools are easy and integrated, but this is not necessarily the case for other CI/CDs.
We asked the panel what they are doing as well as what should be done to secure build pipelines.

\subsection{Secure Strategies / Frameworks}
Participants agreed that threat modeling for the entire build infrastructure is hard.
There are a lot of places where an attacker can change or add that goes straight to the consumer due to automated build systems.
On top of this, a lot of the current defences do not seem to be getting at the heart of the problem.
Strategies such as attestation does not tell you what parts of the build are wrong, it just tells you what is there.
Reproducibility is effective when finding tampering in build code, but compiler tampering is a huge problem that current reproducibility may not be able to handle.

Participants discussed some of their processes for securing their pipeline.
One participant suggested that hermetically sealing builds should prevent build tampering.
Use of Tekton chains for trusted control pipelines and cryptographic provenance (SLSA Build Level 2) helps with creating trails.
Vetting of build tasks before and after they are run.
Some paritipants believed that traceability is as important if not more important than prevention.
Some build pipelines which are believe to be secure like GitHub Actions may not be fully secure and should be looked into.

\subsection{Reproducible Builds}
Reproducibility is an interesting goal to strive for.
It became apparent through discussion that there is no one set definition of reproducibility that everyone follows.
Is it more important that the build can be made twice bit-for-bit or just remade at all?
Participants agreed that the goal of each should be the same and that developers are trying to detect tampering with the system that were not already there in an ordinary build.

One participant put forth that currently reproducibility is not the main problem since 96\% of vulnerable downloads are done on packages that are known to already be vulnerable.
This is not to say that reproducibility is useless.
It does help to capture information about your system and helps to find and isolate elements that get compromised quicker. 
However, this is not something that companies need to fully strive for at the beginning and should be a wishlist item.

\section{Vulnerability Reduction at Scale}

With the use of Static Application Security Testing (SAST) tools comes a lot of vulnerability alerts.
These alerts can be in the thousands for some software products so it becomes imperative that developers are able to respond and handle these vulnerabilities.
One participant stated that 70\% of vulnerabilities are from memory safety issues.
Thus, we asked the panel what strategies have they been employing to reduce vulnerabilities at scale through secure frameworks or moving to memory safe languages.

\subsection{Memory Safe Languages}
While memory safety issues may make up a large majority of the vulnerabilities found in the wild today, moving to memory safe languages does not solve all problems of programming.
One participant stated the phrase "Same bugs, different languages" for the phenomenon they noticed when moving to these memory safe languages.
The same participant stated though that using memory safe languages moving forward is a good practice to do.
Another participant noted that the government would rather use products that use memory safe languages vs ones that do not.
The converse of this was the notion of redeveloping everything is a huge effort as well as battle tested code has its benefits and should not be changed.
If the battle tested code has been pruned for a long time and have not had any issues, it should not be a priority when transitioning code to a safe language.
Participants urged that companies should be changing the old code to the newer languages as needed. 
The goal would be to eventually phase out all of the legacy code, but this does not all need to be done at once.

There are some complications that hold memory safe languages back. 
Memory safe languages such as Rust have a very steep learning curve. 
Finding new developers that know safe languages and practices as well as teaching current developers is a difficult task.
Due to the limitations of some of these languages and the needs of particular software, it might not be possible to make the full switch.
Software life-cycles do not allow for quick updates to entirely new languages and software. 
Phasing out legacy applications to memory safe languages quickly is a very hard endeavour.

\subsection{Secure Practices}
Having secure practices when developing software is important in almost every part of the development cycle. 
Participants agreed that secure practices done early can reduce large vulnerabilities down the line.
One participant discussed their process of checking code before it even gets into the pipeline.
Another participant shared that their frontend protections and sanitation were effective means of reducing vulnerabilities.
Shifting risk such as hosting infrastructure through a third party allows for the company to focus more on whats important to them and not have to worry as much about everything.

If possible, a large team effort to reduce the vulnerabilities over a long period of time was discussed to be effective.
One participant on the panel discussed their success with a dedicated team to recucing thousands of vulnerabilities down to less than 100 in 2.5 years.
They pushed that the team must be dedicated to this task alone as they become faster at solving these issues overtime and can come up with faster ways to solve problems similar in the future.
Automating this process is important, but when considering designing for automation, what to automate first and how fast is the most important.

Once again the theme of speed came up as a secure policy. 
Being able to fix things really quickly rather than having a super heavy framework that slows down production is more beneficial overall.
Potentially contradictory to speed is an understanding needs to be made that code that is written today may be used for a long time so care should be taken to make sure they work correctly in the beginning.
Being able to balance between speed and security is a crucial step that companies need to achieve for their secure policies.

\section{LLMs and Supply Chain}

Recent advancements in AI technology, specifically the rise of LLMs, has lead to widespread use by companies and developers alike.
While this technology is seemingly very useful, it is still in the early stages of its developmental lifestyle. 
New challenges will begin to appear as more and more adopt the use of them into their systems and software.
We asked the panel at this time, how are they leveraging AI/ML into securing the supply chain.

\subsection{Potential Use Cases}
Finding uses for models in non-critical tasks appeared to be the thought process for looking for any kind of task for the LLM, such as threat modeling and translation.
Participants expressed that trying to find other uses for LLMs is hard.
One participant was striving to look into fuzz testing by getting the LLM to generate fuzz tests, but found this hard as they were unsure as to how to go about doing this.
Another participant said they were looking into using LLMs to answer security questions.
Using LLMs as monitors came up again for behavior analysis, anomaly detection in networks, and catching attacks during holidays when attackers are likely to strike.
Use for training developers in security topics such as risk management training as well as helping Blue Teams develop attacks and defenses was also something being considered.

Participants discussed some things that are already being done to use AI including generating environment configurations for builds.
Another participant discussed how they found that LLMs were able to fix 7.5\% of reviewer comments automatically.

The group then discussed that a stretch goal for LLMs is to use them to solve simple vulnerabilities to reduce the number of things that need to be looked at for developers.
They also expressed interest in researching how LLM developers can get the models to generate more secure code and recommend more secure dependencies.

\subsection{Limitations}
While AI offers a lot of benefits, they have a lot of limiting factors that are holding companies back from full adoption.
One of the major issues brought up with LLMs is trust. 
Can companies trust the model to do what they need it to do or will hallucinations be a huge detriment.
When hallucinations happen, it is hard to trace problems back to the source when they do not know what the source is.

One participant made the claim that LLMs will cause bad developers to create huge amounts of bad code.
Small teams that require big efforts will substitute LLMs and will run into a lot of issues when the code does not work well enough. 
An idea brought up was that the liability of LLMs will be a huge factor going forward that is seemingly ignored currently.
Pen testing against LLMs is easy and exploitable (Feels out of place).
The risk from LLMs use itself is poorly understood and highly imminent.

Incorporation into code writing is too much of a risk for proprietary code.
It is possible to be done if you use and maintain your own LLM model with your own training data, but this is very expensive and time consuming to even create the dataset.

When looking specifically at supply chain applications of LLMs, including models into any part of the process will be hard to determine if that model has not been poisoned. 
Verification of safe models is a big question that comes from this.
A direct quote from on the of the participants was "LLMs generally make supply chain security worse".
Companies need to understand better how LLMs work on a fundamental level to make any kind of decision.

\section{Summary}

After discussing all of the topics of the summit, participants were asked what was their key takeaways from all of the talks.

One of the participants expressed that everything that was discussed is hard.
These problems are not fully solvable for any company.
Coming up with solutions that fit the broad scope of every company is impossible.
Another participant stated solutions that can be created can follow the "80-20" rule in which it can get companies started in the right direction and they can prioritize what they care about more.

Security strategies proposed to mitigate the supply chain security issues fell heavily into the defense in depth category.
Being able to secure every part of the pipeline as much as possible will help with overall security 
Engineering effort became the biggest deciding factor in what companies will decide to do over something else.
Cost benefit analysis will be very important in this decision process.
Being able to build software fast seemed to be the most widely accepted sentiment for keeping software as secure as possible.
Priorities of the company will also dictate their policies which will influence how heavy easy defense will need to be.
Knowing the type of product you are trying to produce as well as the organization you work for will help with these decisions.
\section{Acknoledgements}
We would like to thank all of the participants of the Summit. 
All of their discussions and insights will be very beneficial to the open source supply chain security community.
The Summit was organized by Laurie Williams and Alexandros Kapravelos and was recorded by Greg Tystahl.
This material is based upon work supported by the National Science Foundation Grant Nos. 2207008, 2206859, 2206865, and 2206921. 
These grants support the Secure Software Supply Chain Summit (S3C2), consisting of researchers at North Carolina State University, Carnegie Mellon University, University of Maryland, George Washington University, and Paderborn University. 
Any opinions expressed in this material are those of the author(s) and do not necessarily reflect the views of the National Science Foundation.

\bibliographystyle{plain}
\bibliography{s3c2-summit-refs}

\appendix
\section{Full survey questions for panel}
\begin{enumerate}
    \item \textbf{Software Bill of Materials (SBOM):} Where are you in your journey toward producing an SBOM?  Where are you in your journey toward consuming/using the SBOMs of components and products you use?  What challenges have you faced in SBOM production or use and how have you tried to overcome these challenges?  Are you creating a VEX?  How?
    \item \textbf{Vulnerable dependencies:} What process and/or tools do you use to find out that you have a vulnerable dependency? What is your processes for evaluating/prioritizing what dependencies to update and actually updating vulnerable dependencies?   Do you push a new version of a dependency with a major or minor release? 
    \item \textbf{Malicious commits:} How can malicious commits be detected? What do you think signals a suspicious/malicious commit?  What role does the ecosystem play in detecting malicious commits?
    \item \textbf{Build Infrastructure:} What is being done (or should be being done) to secure the build and deploy process/tooling pipeline (a.k.a SLSA practices)?  Are you working toward reproducible builds?
    \item \textbf{Reducing entire classes of vulnerabilities at scale:} Are you moving toward the use of safer languages?  Mandating the use of any secure frameworks?
    \item \textbf{LLMs and Supply Chain Security:} How are you leveraging the recent advances in ML/AI in securing your software supply chain?
\end{enumerate}

\end{document}